\documentclass[longauth,traditabstract]{aa}

\usepackage{graphicx}
\usepackage{txfonts}
\include{epsf}
\newcommand{\be}{\begin{equation}}
\newcommand{\ee}{\end{equation}}
\newcommand{\ba}{\begin{eqnarray}}
\newcommand{\ea}{\end{eqnarray}}

\def\simless{\mathbin{\lower 3pt\hbox
   {$\rlap{\raise 4pt\hbox{$\char'074$}}\mathchar"7218$}}}
\def\simgreat{\mathbin{\lower 3pt\hbox
   {$\rlap{\raise 4pt\hbox{$\char'076$}}\mathchar"7218$}}}

\begin{document}

\title{The Herschel-ATLAS\thanks{Herschel is an ESA space
observatory with science instruments provided by European-led
Principal Investigator consortia and with important participation from
NASA}: Evolution of the 250$\,\mu$m luminosity
function out to $z=0.5$}

\author{
S. Dye\inst{1}
\and
L. Dunne\inst{5}
\and
S. Eales\inst{1}
\and
D.J.B. Smith\inst{5}
\and
A. Amblard\inst{2}
\and
R. Auld\inst{1}
\and
M. Baes\inst{3}
\and 
I.K. Baldry\inst{4}
\and
S. Bamford\inst{5}
\and
A.W. Blain\inst{31}
\and
D.G. Bonfield\inst{6}
\and
M. Bremer\inst{33}
\and
D. Burgarella\inst{7}
\and
S. Buttiglione\inst{8}
\and
E. Cameron\inst{9}
\and
A. Cava\inst{10}
\and
D.L. Clements\inst{11}
\and
A. Cooray\inst{2}
\and
S. Croom\inst{12}
\and
A. Dariush\inst{1}
\and
G. de Zotti\inst{8}
\and
S. Driver\inst{13}
\and
J.S. Dunlop\inst{24}
\and
D. Frayer\inst{14}
\and
J. Fritz\inst{3}
\and
Jonathan P. Gardner\inst{30}
\and
H.L. Gomez\inst{1}
\and
J. Gonzalez-Nuevo\inst{15}
\and
D. Herranz\inst{16}
\and
D. Hill\inst{13}
\and
A. Hopkins\inst{17}
\and
E. Ibar\inst{18}
\and
R.J. Ivison\inst{18}
\and
M.J. Jarvis\inst{6}
\and
D.H. Jones\inst{17}
\and
L. Kelvin\inst{13}
\and
G. Lagache\inst{19}
\and
L. Leeuw\inst{20}
\and
J. Liske\inst{21}
\and
M. Lopez-Caniego\inst{16}
\and
J. Loveday\inst{22}
\and
S. Maddox\inst{5}
\and
M.J. Micha{\l}owski\inst{24}
\and
M. Negrello\inst{23}
\and
P. Norberg\inst{24}
\and
M.J. Page\inst{32}
\and
H. Parkinson\inst{24}
\and
E. Pascale\inst{1}
\and
J.A. Peacock\inst{24}
\and
M. Pohlen\inst{1}
\and
C. Popescu\inst{25}
\and
M. Prescott\inst{4}
\and
D. Rigopoulou\inst{28}
\and
A. Robotham\inst{13}
\and
E. Rigby\inst{5}
\and
G. Rodighiero\inst{8}
\and
S. Samui\inst{15}
\and
D. Scott\inst{34}
\and
S. Serjeant\inst{23}
\and
R. Sharp\inst{17}
\and
B. Sibthorpe\inst{18}
\and
P. Temi\inst{20}
\and
M.A. Thompson\inst{6}
\and 
R. Tuffs\inst{26}
\and
I. Valtchanov\inst{27}
\and
P.P. van der Werf\inst{29}
\and
E. van Kampen\inst{21}
\and
A. Verma\inst{28}
}

\institute{
School of Physics and Astronomy, Cardiff University, Queens Buildings,
The Parade, Cardiff, CF24 3AA, UK \\
\email{s.dye@astro.cf.ac.uk} 
\and
Dept. of Physics \& Astronomy, University of California, Irvine, 
CA 92697, USA
\and
Sterrenkundig Observatorium, Universiteit Gent, Krijgslaan 281 S9,
B-9000 Gent, Belgium
\and
Astrophysics Research Inst., Liverpool John Moores University,
12 Quays House, Egerton Wharf, Birkenhead, CH41 1LD, UK
\and
School of Physics and Astronomy, University of Nottingham,
University Park, Nottingham NG7 2RD, UK
\and
Centre for Astrophysics Research, Science and Technology Research
Institute, University of Hertfordshire, Herts AL10 9AB, UK
\and
Laboratoire d'Astrophysique de Marseille, UMR6110 CNRS, 38 rue F.
Joliot-Curie, F-13388 Marseille France
\and
University of Padova, Department of Astronomy, Vicolo Osservatorio
3, I-35122 Padova, Italy
\and
ETH Zurich, Insitute for Astronomy, HIT J12.3, CH-8093 Zurich, Switzerland
\and
Instituto de Astrof\'{i}sica de Canarias (IAC) and Departamento de
Astrof\'{i}sica, Universidad de La Laguna (ULL), La Laguna, Tenerife, Spain
\and
Astrophysics Group, Imperial College, Blackett Laboratory, Prince
Consort Road, London SW7 2AZ, UK
\and
Sydney Institute for Astronomy, School of Physics, University of
Sydney, NSW 2006, Australia
\and
SUPA, School of Physics and Astronomy, University of St. Andrews, 
North Haugh, St. Andrews, KY16 9SS, UK
\and
National Radio Astronomy Observatory,  PO Box 2, Green Bank, WV 24944, USA
\and
Scuola Internazionale Superiore di Studi Avanzati, via Beirut 2-4,
34151 Triest, Italy
\and
Instituto de F\'isica de Cantabria (CSIC-UC), Santander, 39005, Spain
\and
Anglo-Australian Observatory, PO Box 296, Epping, NSW 1710, Australia
\and
UK Astronomy Technology Center, Royal Observatory, 
Edinburgh, EH9 3HJ, UK
\and
Institut d'Astrophysique Spatiale, B\^{a}timent 121, F-91405 Orsay, France; 
Universit\'{e} Paris-Sud 11 and CNRS (UMR 8617), France
\and
Astrophysics Branch, NASA Ames Research Center, Mail Stop 245-6, 
Moffett Field, CA 94035, USA
\and
European Southern Observatory, Karl-Schwarzschild-Strasse 2 D-85748,
Garching bei Munchen, Germany
\and
Astronomy Centre, Department of Physics and Astronomy, School
of Maths and Physical Sciences, Pevensey II Building, University of
Sussex, Falmer, Brighton, BN1 9QH, UK
\and
Dept. of Physics and Astronomy, The Open University, Milton Keynes, 
MK7 6AA, UK
\and
SUPA, Institute for Astronomy, University of Edinburgh, Royal Observatory,
Blackford Hill, Edinburgh EH9 3HJ, UK 
\and
Jeremiah Horrocks Institute, University of Central Lancashire,
Preston PR1 2HE, UK
\and
Max Planck Institute for Nuclear Astrophysics (MPIK), Saupfercheckweg 1,
69117 Heidelberg, Germany
\and
Herschel Science Centre, ESAC, ESA, PO Box 78, Villanueva de la
Ca\~nada, 28691 Madrid, Spain
\and
Oxford Astrophysics, Denys Wilkinson  Building, University of Oxford, 
Keble Road, Oxford, OX1 3RH
\and
Leiden Observatory, Leiden University, P.O. Box 9513, NL - 2300 RA Leiden,
The Netherlands
\and
Astrophysics Science Division, Observational Cosmology Laboratory, 
Code 665, Goddard Space Flight Center, Greenbelt, MD 20771, USA
\and
Caltech, MS247-19, Pasadena, CA 91125, USA
\and
University College London, Mullard Space Science Laboratory,
Holmbury St. Mary, Dorking, Surrey, RH5 6NT, UK
\and
H H Wills Physics Laboratory, University of Bristol,
Tyndall Avenue, Bristol, BS8 1TL, UK
\and
Department of Physics \& Astronomy, University of British Columbia,
Vancouver, BC, V6T 1Z1, Canada
}

\titlerunning{The H-ATLAS 250$\,\mu$m luminosity function 
out to $z=0.5$}

\date{Received 31 March 2010}

\abstract{We have determined the luminosity function of
250$\,\mu$m-selected galaxies detected in the $\sim$14\,deg$^2$
science demonstration region of the Herschel-ATLAS project out to a
redshift of $z=0.5$. Our findings very clearly show that the
luminosity function evolves steadily out to this redshift. By
selecting a sub-group of sources within a fixed luminosity interval
where incompleteness effects are minimal, we have measured a smooth
increase in the comoving 250$\,\mu$m luminosity density out
to $z=0.2$ where it is $3.6^{+1.4}_{-0.9}$ times higher than the
local value.}


\maketitle


\section{Introduction}

Measurement of the galaxy luminosity function (LF) constitutes one of
the most fundamental statistical constraints that can be placed on
models of galaxy formation, and hence the build up of large scale
structure, in the universe. Since half of the energy ever emitted by
galaxies has been absorbed by dust and re-radiated at far-infrared and
sub-millimetre (submm) wavelengths (Fixsen et al. \cite{fixsen98})
and, because knowledge of the statistical properties of submm sources is
relatively sparse, determination of the submm LF provides a crucial
missing piece in a fully comprehensive model of galaxy evolution.

Following their detection in the first deep submm and mm surveys
(e.g., Smail et al. \cite{smail97}; Hughes et al. \cite{hughes98};
Eales et al. \cite{eales99}; Bertoldi et al. \cite{bertoldi00}), much
has been learned about the dusty high-redshift sources selected at
such wavelengths. Although many studies have argued that these sources
are likely ancestors of local ellipticals (e.g., Scott et
al. \cite{scott02}; Dunne et al. \cite{dunne03}) little progress in
verifying this assertion has been made since.  The main reason for
this is the preponderance of high redshift submm/mm-selected sources,
owing to the strong negative k-correction and small survey areas. The
resulting low numbers of sources at redshifts $z<1$ has therefore
precluded evolutionary studies over the last $\sim$60\% of the
Universe's history.

In particular, despite the local submm LF being first determined a
decade ago (Dunne et al. \cite{dunne00}), little has been added to our
comprehension of how the LF has evolved over the last $\sim 7$\,Gyr,
until very recently. Observations conducted using the Balloon-borne
Large Aperture Submm Telescope (BLAST; Devlin et al. \cite{devlin09})
have made significant improvements with much enhanced sensitivity to
$z<1$ sources. As a result, direct estimates of the LF at 250, 350 and
500$\,\mu$m were made by Eales et al.  (\cite{eales09}) who detected
strong evolution, particularly among the higher luminosity systems,
from $z=1$ to the present day. However, the accuracy of these findings
is limited by the small number of sources used ($\sim 50$ at $z<0.5$)
and source confusion due to the angular resolution of BLAST.

In this letter, we present our measurement of the LF of
250$\,\mu$m-selected galaxies detected by the Herschel Space
Observatory ({\sl Herschel}; Pilbratt et al. \cite{pilbratt10}) over a
$\sim 14\,$deg$^2$ region acquired as part of the science demonstration
observations of the {\sl Herschel}-Astrophysical Terahertz Large Area
Survey (H-ATLAS; Eales et al. \cite{eales10}). These data offer a
significant improvement over the BLAST data in terms of their
increased sensitivity, higher angular resolution and greater areal
coverage, resulting in $\sim 20$ times the number of sources
with which to compute the LF.

Throughout this letter, the following cosmological parameters have
been assumed; ${\rm H}_0=71\,{\rm km\,s}^{-1}\,{\rm Mpc}^{-1}$,
$\Omega_{\rm m}=0.27$, $\Omega_{\Lambda}=0.73$.

\section{Data}

The $4^\circ \times 4^\circ$ H-ATLAS science demonstration field was
observed with the Spectral and Photometric Imaging Receiver (SPIRE;
Griffin et al. \cite{griffin10}) at the wavelengths 250, 350 and
500$\,\mu$m and with the Photodetector Array Camera and Spectrometer
(PACS; Poglitsch et al. \cite{poglitsch10}) at 100 and 160$\,\mu$m.
The field, centred at the co-ordinates ($09^h 05^m 30^s, +00^\circ 30'
00''$), was scanned twice in parallel mode.  The $5 \sigma$ point
source sensitivities of the resulting beam-convolved maps, including
confusion noise, are 132, 126, 32, 36 and 45\,mJy and the beam
sizes expressed as full width at half maximum (FWHM) are $9''$,
$13''$, $18''$, $25''$ and $35''$ at 100, 160, 250, 350 and
500$\,\mu$m respectively.  Details of the SPIRE and PACS map-making
are given in Pascale et al. (in preparation) and Ibar et al. (in
preparation) respectively.

Sources were initially extracted from the 250$\,\mu$m noise-weighted
beam-convolved map in a central 14.4\,deg$^2$ region above a
significance of 2.5$\sigma$. For each source, 350 and 500$\,\mu$m
fluxes were then estimated from the appropriate beam-convolved map at
positions determined at 250$\,\mu$m. Extended source fluxes were
measured in apertures matched to identified (IDed) optical counterpart
sizes (see below). 6878 sources were detected with a significance of
$\geq 5\sigma$ in any one band.  Fluxes at 100 and 160$\,\mu$m were
assigned by matching to $\geq 3 \sigma$ PACS sources within a
positional tolerance of $10''$. The analysis described hereafter
applies to the 6613 sources detected at $\geq 5 \sigma$ at
250$\,\mu$m. Full details of the source extraction are given in Rigby
et al. (in preparation).

Since our field (and H-ATLAS at large) is lacking in the radio
and mid-infrared data traditionally used to identify counterparts to
submm sources, we have taken a different approach and matched directly
to optical counterparts.  It is possible to attain a reasonable rate
of secure optical IDs in this way with H-ATLAS because, unlike
existing submm surveys at 850$\,\mu$m-1.1\,mm sensitive to high
redshift sources ($z_{\rm median} \simeq 2.5$), H-ATLAS sources lie at
substantially lower redshifts ($z_{\rm median} \simless 1.0$) on
average. This means that we can use shallow optical imaging, where a
low surface number density of sources allows for much less ambiguous
IDs.

We used the likelihood ratio (LR) method of Sutherland \& Saunders
(\cite{sutherland92}) to perform the matching, which uses the submm
positional uncertainties and the magnitude distribution of
counterparts to assign the likelihood that a particular optical source
is physically associated with a target submm galaxy. We searched for
optical counterparts at $r\leq 22.4$ from the Sloan Digital Sky Survey
(SDSS) seventh data release (Abazajian et al. \cite{abazajian09})
within $10"$ of every 250$\,\mu$m SPIRE source. The LR technique
assigns a reliability parameter, R$_{\rm LR}$, to each match, which
indicates the probability that the counterpart is the correct ID. The
calculation of R$_{\rm LR}$ includes the probability that the true
counterpart may be below the detection limit of the survey and
accounts for other counterparts within the same search radius. To
remove unreliable counterparts, all those with R$_{\rm LR}< 80\%$ were
discarded, leaving a total of 2267 submm sources with unique optical
counterparts. We refer the reader to Smith et al. (in preparation) for
an exhaustive account of the ID procedure.

\begin{figure}
\epsfxsize=65mm
{\hfill
\epsfbox{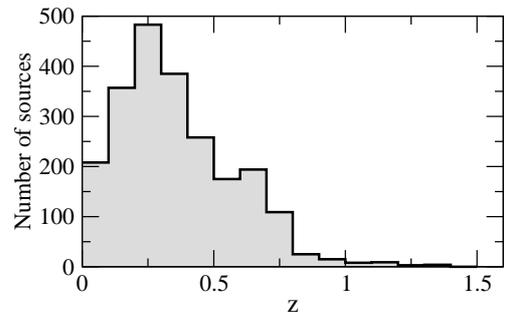}
\hfill}
\caption{Redshift distribution of the 2241 sources that are
detected at 250$\,\mu$m with $\geq 5 \sigma$ significance, and
have an optical counterpart with a redshift and a reliability 
of association of $\geq 80$\%. The 1688 of these at $z\leq 0.5$
form the sample analysed in this letter.}
\label{zhist}
\end{figure}

Of these 2267 counterparts, 876 have spectroscopic redshifts (spec-zs)
acquired either by the SDSS, the Galaxy And Mass Assembly survey
(Driver et al. \cite{driver09}), the 2dF redshift survey (Colless et
al. \cite{colless01}) or the 6dF redshift survey (Jones et
al. \cite{jones09}).  For the remaining counterparts, photometric
redshifts (photo-zs) were estimated by applying the ANNz neural
network code (Collister \& Lahav \cite{collister04}) to the SDSS
optical photometry and also near-infrared photometry taken from the
seventh data release of the UKIRT Infrared Deep Sky Survey (UKIDSS;
Lawrence et al. 2007). Details of these photo-zs are given in Smith et
al. (in preparation). A total of 2239 sources were assigned
photo-zs, with a total of 2241 sources having a redshift of either
type. Figure \ref{zhist} shows the redshift distribution of these 2241
sources. The 1688 sources at $z\leq 0.5$ form the sample to which we
apply our analysis in this letter. In every case, we used a spec-z in
preference to a photo-z, although there is excellent agreement between
the two, with a standard deviation of $(z_{\rm phot} - z_{\rm
spec})/(1 + z_{\rm spec})=0.039$ over the sample.

\section{The luminosity function}

\subsection{SED fitting}

A modified black-body spectral energy distribution (SED) was fitted to
the SPIRE photometry, and, where available (272 sources), PACS
photometry, for each source. In the fitting, we allowed the rest-frame
dust temperature, T, to vary between $10{\rm K < T < 50 K}$, we fixed
the dust emissivity index to $\beta=1.5$ and we fixed the redshift to
either the photo-z, or, preferentially, when available, the
spec-z. The temperature of 342 sources (none with PACS photometry)
could not be reliably constrained. For these, we re-fitted the SED,
fixing the temperature to the median of the sample (see below).

Excluding the sources where T was fixed, we found a median dust
temperature across the whole sample of 26K with a standard deviation
of 8K (or $23\pm7$\,K for $\beta=2$ ).  This result closely agrees
with the rest-frame dust temperatures determined for BLAST sources
measured by Dye et al. (\cite{dye09}) as well as those of the sample
of H-ATLAS galaxies studied in Amblard et al. (\cite{amblard10}; see
also other temperature comparisons therein).

For each source, we measured the rest-frame 250$\,\mu$m luminosity,
L$_{250}$, by integrating the rest-frame SED over the 250$\,\mu$m
SPIRE bandpass function. Errors on L$_{250}$ were determined by
propagating redshift errors and temperature errors obtained from the
SED fit.

\subsection{Estimating the LF}
\label{sec_est_lf}

We based our measurement of the LF on the estimator
\be
\phi=\Sigma_i(V_{{\rm max},i})^{-1}  \, ,
\ee 
where $V_{{\rm max},i}$ is the comoving volume out to the maximum
redshift that source $i$ could be placed and remain above the
250$\,\mu$m {\em and} optical $r$-band detection threshold.  We used
an Sb type SED for the optical k-correction and the SED fit as
described in the previous section for the k-correction at 250$\,\mu$m.
The sum here acts over all sources in a given luminosity and redshift
bin. We computed the total error on $\phi$ as the quadrature sum of
its formal error, $\sqrt{\Sigma_i(V_{{\rm max},i})^{-2}}$, and the
standard deviation of the scatter measured in each bin in performing a
Monte Carlo simulation in which redshifts and photometry were
randomised according to their errors.

\begin{figure}
\epsfxsize=65mm
{\hfill
\epsfbox{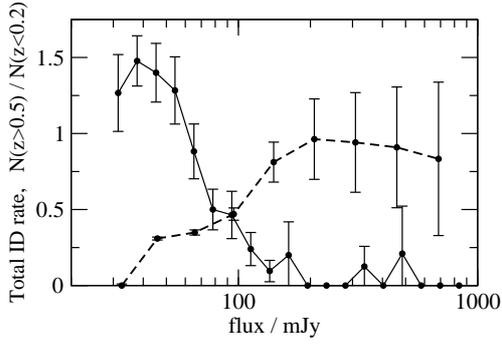}
\hfill}
\caption{Rate of identification of optical counterparts to 250$\,\mu$m
sources (dashed line) and ratio of the number of $z>0.5$ sources to
the number of $z<0.2$ sources (continuous line), as a function of
250$\,\mu$m flux. Both quantities were determined using the sample of
2241 counterparts. Poisson errors are plotted.}
\label{completeness}
\end{figure}

An important consideration when computing the LF is incompleteness.
There are two types of incompleteness at play in our case.  Firstly,
250$\,\mu$m sources with optical fluxes lower than the $r$-band
detection threshold are missing.  This is a stronger effect at higher
redshifts and lower luminosities.  Figure \ref{completeness} shows how
the ID rate (i.e., identifying with an optical counterpart with
$R_{\rm LR}>80\%$) varies with 250$\,\mu$m flux over the full sample
of 6613 sources detected at 250$\,\mu$m.  The figure also shows that
the ratio of the number of $z>0.5$ sources to the number of $z<0.2$
sources\footnote{These two redshift limits were chosen to obtain
approximately equally sized sub-samples.} increases rapidly towards
low 250$\,\mu$m fluxes. Together, these two facts imply that the
unidentified 250$\,\mu$m sources predominantly lie at high redshifts
where their optical flux falls beneath the sensitivity limit of the
counterpart catalogue.

However, the fraction of missing sources at $z<0.5$ (where we have
computed the LF -- see Section \ref{sec_lf_evol}) with undetected
optical flux must still be quantified and accounted for. We therefore
plotted the optical $r$-band flux against 250$\,\mu$m flux for sources
in different redshift slices and found a clear positive correlation in
every slice. At redshifts $z\simgreat 0.2$, the faint end of the locus
of points on this plot becomes cut by the optical sensitivity
limit. Using this result, we were able to determine a 250$\,\mu$m flux
limit for each redshift slice where we estimate that $<5\%$ of sources
are missing due to the optical sensitivity. We then limited our
computation of the LF in each redshift bin by the corresponding
250$\,\mu$m luminosity limit, ensuring $>95\%$ completeness at all
redshifts. H-ATLAS completeness effects will be discussed at length in
forthcoming work, including Rigby et al. (in preparation).

The second incompleteness effect is a consequence of the reliability cut.
Approximately one half of the 250$\,\mu$m sources that were
matched to an optical source, were rejected by applying a minimum
reliability of 80\%. A certain fraction of these will be genuine
counterparts. To estimate this fraction, we applied a correction
factor. This factor was computed by reversing the algorithm used in
the ID procedure (see Smith et al., in preparation) that determines a
counterpart's reliability from the radial offset between the submm and
optical positions, the optical $r$-band magnitude and the
signal-to-noise ratio of the submm source. We therefore computed the
submm-optical offset, $r_{80}$, corresponding to a reliability of 80\%
for the full range of combinations of 250$\,\mu$m source
signal-to-noise ratio and $r$-band magnitude seen in the data.  For
each combination, the correction factor, $c$, was then calculated as
the reciprocal of the fraction of counterparts that would be IDed
within an offset of the smaller of $r_{80}$ and the ID search radius
of $10''$, i.e.,
\be
c = A\left(\int_0^{{\rm min}(r_{80},10'')}
r \exp\left(-\frac{r^2}{2\sigma^2}\right) {\rm d}r \right)^{-1},
\ee
where the normalisation $A$ is set such that one would obtain $c=1$
were the integral evaluated between 0 and $\infty$.  Here,
$\sigma$ depends on the submm signal-to-noise, $\mu$, and the beam
FWHM according to $\sigma=0.6*{\rm FWHM}/\mu$, as given by Ivison et
al. (\cite{ivison07}). At 250$\,\mu$m, $\sigma$ therefore varies from
$\sim 2.2''$ for a source with $\mu=5$ to $\sim 1.1''$ for a source
with $\mu=10$. These values are consistent with the distribution of
offsets obtained in matching to the SDSS. The minimum value of
$\sigma$ was limited to $1.0''$, to account for the SPIRE pointing
error and map pixel size (see Smith et al., in preparation, for more
details).

The resulting correction factor was then applied by modifying the LF
estimator to $\Sigma_i\,c_i(V_{{\rm max},i})^{-1}$ where $c_i$ is the
correction factor corresponding to the $i$th source's 250$\,\mu$m
signal-to-noise and $r$ band counterpart magnitude.  The average
correction factor, weighted by the counterpart number counts, ranges
from $\sim 1.2$ for $5 \sigma$ 250$\,\mu$m sources, through $\sim 1.1$
for $8 \sigma$ sources to $\sim 1.0$ for $10 \sigma$ sources.

Finally, we estimated the expected number of false counterparts by
summing the quantity $1-R_{\rm LR}$. Within the 1688 sources at $z\leq
0.5$ used for computation of our LF, we estimate a total of 81 false
counterparts. This false ID rate of $\sim 5\%$ shows no noticable
correlation with redshift or luminosity and little variation
between LF bins across the redshift-luminosity plane. We treated
the false counterpart rate as an additional source of error and added
it in quadrature to the error on each LF bin.

\section{Results}

\subsection{LF evolution}
\label{sec_lf_evol}

Using the methods outlined previously, we determined the LF in five
redshift bins of width $\Delta z=0.1$ from $z=0$ to $z=0.5$. The total
error budget includes the formal error on the LF estimator that
accounts for Poisson noise, false IDs and the scatter measured in the
Monte Carlo simulation which randomises redshift and photometry.

\begin{figure*}
\epsfxsize=115mm
{\hfill
\epsfbox{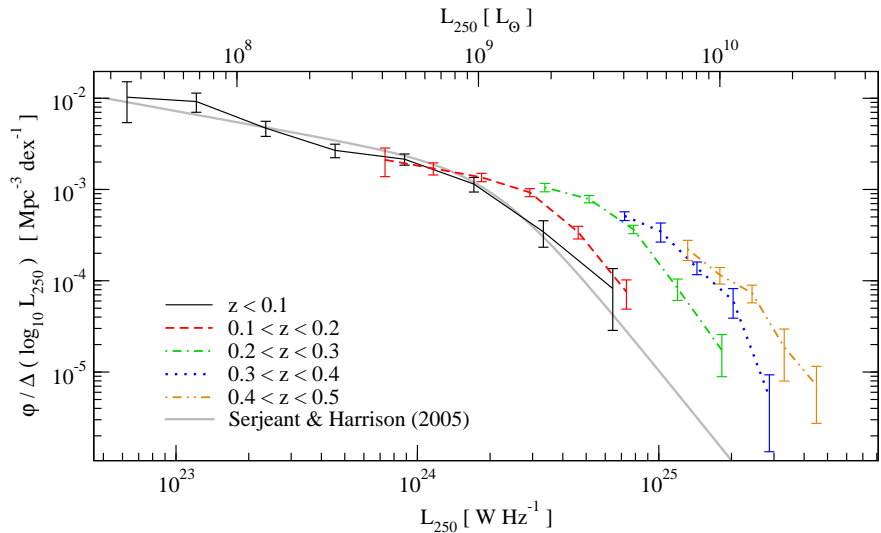}
\hfill}
\caption{The 250$\,\mu$m LF in five different redshift bins of width
$\Delta z=0.1$ from $z=0$ to $z=0.5$. Errors account for Poisson
noise, redshift and photometric errors and the expected false ID
rate. The thicker grey line shows the local 250$\,\mu$m LF predicted
by Serjeant \& Harrison (\cite{serjeant05}).}
\label{lf}
\end{figure*}

Figure \ref{lf} shows the LF in each of the five redshift bins.  The
figure clearly shows that the LF exhibits significant evolution out to
$z=0.5$. At a given luminosity, the comoving space density increases
steadily with redshift. This is consistent with the findings of Eales
et al.  (\cite{eales09}), although our detection of evolution is
considerably more significant. The figure also shows that the local
250$\,\mu$m LF predicted by Serjeant \& Harrison (\cite{serjeant05})
agrees very well with our $z<0.1$ LF.

In each redshift bin, the slope of the LF becomes shallower toward
lower luminosities. Our allowed incompleteness of up to 5\% is
insufficient to account for the magnitude of this effect, although it
is possible that there may be additional mild incompleteness in
the source extraction process at low 250$\,\mu$m fluxes. This will be
quantified in a later paper once the source extraction has been
formally characterised (Rigby et al., in preparation). 

\subsection{Luminosity density evolution}

Having established significant evolution of the LF, we investigated
evolution of the comoving luminosity density. Clearly, incompleteness
will preclude an accurate measurement out to any significant redshift.
However, by limiting the calculation to a sub-group of sources within
a fixed luminosity interval where incompleteness is small, it is
possible to estimate the strength of evolution in the sub-group over a
larger redshift range.

\begin{figure}
\epsfxsize=70mm
{\hfill
\epsfbox{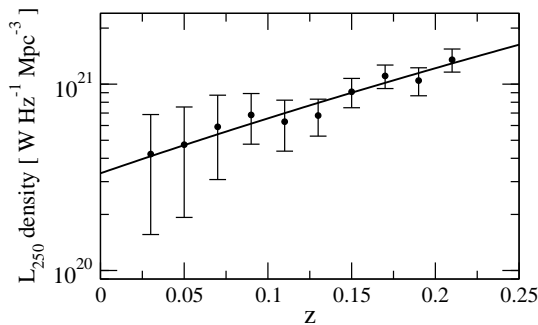}
\hfill}
\caption{Evolution of comoving rest-frame 250$\,\mu$m luminosity
density out to $z\simeq 0.2$ for sources with luminosity in the range
${\rm 10^{9} < L_{250}/L_\odot < 5\times 10^{9}}$. Errors were determined
using a Monte Carlo simulation which randomised redshifts and photometry
and account for Poisson noise. The line is the fit $(1+z)^{7.1}$.}
\label{lum_dens}
\end{figure}

We therefore computed the comoving rest-frame 250$\,\mu$m luminosity
density of a sub-group of sources with luminosities in the range ${\rm
10^{9} < L_{250}/L_\odot < 5\times 10^{9}}$, allowing measurement
up to $z \simeq 0.2$ before incompleteness becomes significant.
We note that within this luminosity and redshift range, approximately
85\% of the sources have spectroscopic redshifts with negligible errors.
Figure \ref{lum_dens} shows the results. The luminosity
density exhibits a steady and significant increase with redshift,
which when fit with the form $(1+z)^n$ yields a value of
$n=7.1^{+2.1}_{-1.4}$. This corresponds to an increase in luminosity
density by a factor of $3.6^{+1.4}_{-0.9}$ from the present day to a 
redshift of $z=0.2$. This scaling is consistent with Saunders et al. 
(\cite{saunders90}) who measured a density scaling of
$(1+z)^{6.7\pm2.3}$ out to $z=0.25$ for 60$\,\mu$m-selected galaxies
but stronger than the scaling $(1+z)^{3.9\pm0.7}$ 
measured by Le Floc'h et al. (\cite{lefloch05}) to $z\sim 1$ for
24$\,\mu$m-selected galaxies.

\section{Conclusion}

One of the key goals of H-ATLAS will be to understand the nature of
the evolution detected in this letter. In turn, we aim to improve our
understanding of the evolutionary link between high redshift and local
submm systems.

This letter has only considered sources selected at 250$\,\mu$m,
merely one of the five wavebands on offer from H-ATLAS.
Furthermore, the 14\,deg$^2$ of survey data analysed in this work
represent only 2.5\% of the final, proposed H-ATLAS survey area. A
repeat of the analysis presented here with the final survey data,
would therefore result in the quoted uncertainties falling by at least
a factor of five. In light of these considerations, it is clear that
our results offer only a small glimpse of the anticipated wealth of
science that H-ATLAS has to offer.


\begin{acknowledgements}

SD Acknowledges the UK STFC for support.

\end{acknowledgements}


\end{document}